\documentclass[10pt,twocolumn,letterpaper]{article}

\usepackage[pagenumbers]{wacv} 

\usepackage{graphicx}
\usepackage{amsmath}
\usepackage{amssymb}
\usepackage{booktabs}
\usepackage{wacv}
\usepackage{times}
\usepackage{epsfig}
\usepackage{pifont}
\usepackage{diagbox}
\usepackage{tabularx}
\usepackage{multirow}
\usepackage{blindtext}
\usepackage{enumitem}
\newcommand{\jhk}[2]{\textcolor{black}{{}#2}}

\usepackage[pagebackref,breaklinks,colorlinks]{hyperref}

\usepackage[capitalize]{cleveref}
\crefname{section}{Sec.}{Secs.}
\Crefname{section}{Section}{Sections}
\Crefname{table}{Table}{Tables}
\crefname{table}{Tab.}{Tabs.}

\makeatletter
\def\@fnsymbol#1{%
   \ifcase#1\or
   \TextOrMath \textdagger \dagger\or
   \TextOrMath \textdaggerdbl \ddagger \or
   \TextOrMath \textsection  \mathsection\or
   \TextOrMath \textparagraph \mathparagraph\or
   \TextOrMath \textbardbl \|\or
   \TextOrMath {\textdagger\textdagger}{\dagger\dagger}\or
   \TextOrMath {\textdaggerdbl\textdaggerdbl}{\ddagger\ddagger}\else
   \@ctrerr \fi
}
\makeatother

\begin{document}
\title{Adaptive Latent Diffusion Model for 3D Medical Image to Image Translation: Multi-modal Magnetic Resonance Imaging Study}
\author{
    Jonghun Kim, \quad
    Hyunjin Park \thanks{Corresponding Author} \\
    Department of Electrical and Computer Engineering \\
    Sungkyunkwan University, Suwon, Korea \\
    {\tt\small iproj2@g.skku.edu, hyunjinp@skku.edu}
}

\maketitle
\thispagestyle{empty}

\begin{abstract}
   Multi-modal images play a crucial role in comprehensive evaluations in medical image analysis providing complementary information for identifying clinically important biomarkers. However, in clinical practice, acquiring multiple modalities can be challenging due to reasons such as scan cost, limited scan time, and safety considerations.  In this paper, we propose a model based on the latent diffusion model (LDM) that leverages switchable blocks for image-to-image translation in 3D medical images without patch cropping.  The 3D LDM combined with conditioning using the target modality allows generating high-quality target modality in 3D overcoming the shortcoming of the missing out-of-slice information in 2D generation methods. The switchable block, noted as multiple switchable spatially adaptive normalization (MS-SPADE), dynamically transforms source latents to the desired style of the target latents to help with the diffusion process. The MS-SPADE block allows us to have one single model to tackle many translation tasks of one source modality to various targets removing the need for many translation models for different scenarios. \jhk{}{Our model exhibited successful image synthesis across different source-target modality scenarios and surpassed other models in quantitative evaluations tested on multi-modal brain magnetic resonance imaging datasets of four different modalities and an independent IXI dataset.  Our model demonstrated successful image synthesis across various modalities even allowing for one-to-many modality translations. Furthermore, it outperformed other one-to-one translation models in quantitative evaluations. Our code is available at \url{https://github.com/jongdory/ALDM/}}
\end{abstract}

\begin{figure} [t]
    \includegraphics[width=\columnwidth]{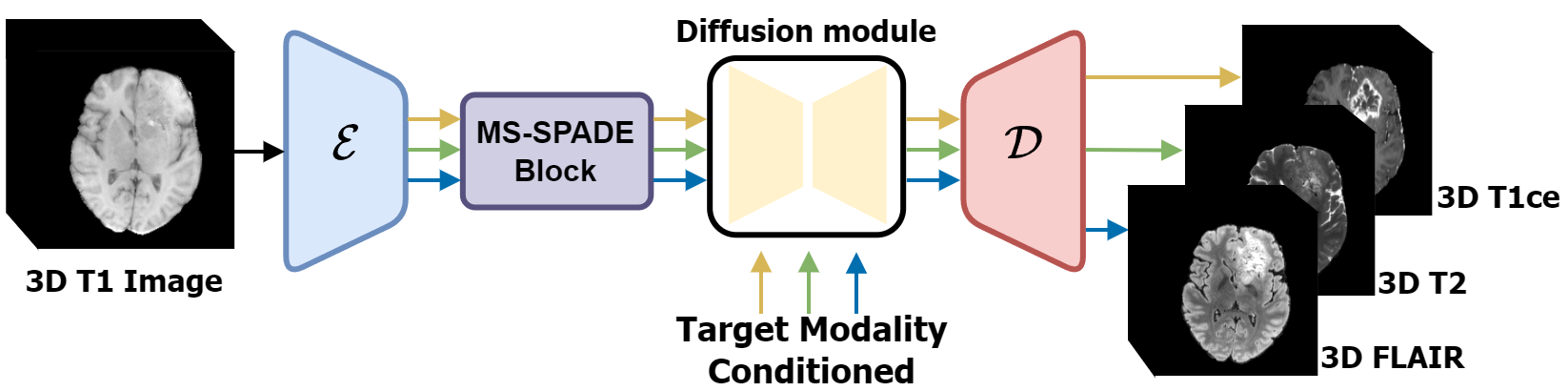}
    \caption{Overview of the proposed image-to-image translation process based on latent diffusion model. Our model utilizes the proposed MS-SPADE to transform the latent representation into the target latents and enables image synthesis in the desired target modality through conditioning.} 
    \label{fig1}
    \vspace{-12pt}
\end{figure}

\section{Introduction}

Multi-modal images are crucial for a comprehensive evaluation in medical image analysis. To identify clinically important biomarkers, multi-modal 3D images with complementary information are often required. Multi-modal imaging serves as a method to compensate for the limitations of individual imaging techniques and enables precise examination \cite{liu2018digital, shen2019brain, tong2017multi}. \jhk{}{Medical image analysis, especially diagnosis of brain tumors, is typically performed using magnetic resonance imaging (MRI) and various MRI techniques such as 3D T1-weighted, T1 contrast-enhanced weighted, T2-weighted, and FLAIR are used, because the distinct modalities (T1, FLAIR, etc.) are different imaging sequences that provide varying contrasts and provide complementary information (e.g., edema, enhancing tumor, and necrosis/non-enhancing tumor) about brain tumor that appears over multiple consecutive slices requiring 3D approach. \cite{menze2014multimodal, bakas2017advancing,bakas2017segmentation}. T1 is sensitive to acute bleeding and clearly displays contrast agents, helping identify vascular structures, tumors, and inflammation areas. T2 highlight conditions like edema or inflammation. FLAIR suppresses cerebrospinal fluid signals and is useful for detecting pathologies such as multiple sclerosis lesions \cite{menze2014multimodal, bakas2017advancing, patel2017t2, shukla2017advanced}.
This is an important application of computer vision in medicine.} However, in clinical settings, obtaining multiple modalities can be challenging due to factors such as scan costs, limited scan time, and safety considerations. Consequently, certain modalities may be missing. This absence of modalities can have a detrimental impact on the quality of diagnosis and treatment. Furthermore, deep learning models that rely on multi-modal data also suffer from reduced performance when crucial modalities are unavailable for training.

Many effective deep learning methods have been proposed to address the information gap caused by missing modalities \cite{van2015does, li2014deep, havaei2016hemis, chen2019robust, zhou2020brain, NEURIPS2021_0f281810, dalmaz2022resvit}. One approach is to train a single model that learns the shared representation of multi-modalities to handle various missing scenarios \cite{van2015does, li2014deep, chen2019robust, zhou2020brain}. For example, with three missing modalities out of four complete MRI modalities, we need to consider 12 ($3\times4$) missing scenarios. The number of scenarios grows quickly as the number of modalities grows.  However, these methods are difficult to generalize across diverse scenarios and are limited to specific tasks, thus not providing a fundamental solution.

Another approach is to generate missing modality images using generative models. Among the existing methods for generating missing modalities, generative adversarial models (GANs) \cite{NIPS2014_5ca3e9b1, isola2017image, zhu2017unpaired, chen2020reusing, NEURIPS2021_0f281810, dalmaz2022resvit} have been used for medical image-to-image translation. However, these methods are 2D and thus  do not consider the information from adjacent slices in 3D medical images. They generate slices independently making it difficult to ensure consistent information between adjacent slices. 3D methods for handling medical imaging data are computationally expensive and challenging to apply. To address this issue, a common approach is to train the model using patches during the training stage and generate results using a sliding window method during the inference stage. However, these methods may result in the loss of global information and a potential decrease in image translation performance. Furthermore, existing methods typically generate a given target modality from one or many source(s) (i.e., one-to-one or many-to-one) \cite{9004544, NEURIPS2021_0f281810, dalmaz2022resvit}. To generate multiple targets from a single source modality, these methods require a separate model for each target resulting in increased complexity. Therefore, there is a need for a model that can perform image translation using 3D medical imaging without patch cropping and allow generating flexible targets from a single source modality.

In recent years, the diffusion model (DM) has emerged as a powerful generative model with high-quality generation capabilities, positioning itself as a potential alternative to  GANs \cite{ho2020denoising, dhariwal2021diffusion, saharia2022palette, rombach2022high, pinaya2022brain, croitoru2023diffusion}. The diffusion model leverages cross-attention and flexible conditioning to enable the generation of desired images. Additionally, the latent diffusion m odel (LDM) defines a denoising task in the compressed latent space of an image significantly reducing computational costs while demonstrating the ability to generate high-quality images at a lower cost \cite{rombach2022high}. LDM demonstrated its applicability to 3D medical images without the need for patch cropping. This advancement allows for the application of LDM to medical images even in their native 3D form \cite{khader2023denoising, pinaya2022brain}.

In this study, we propose a model that utilizes the conditioning mechanism of LDM to perform image-to-image translation in 3D medical images without patch cropping maintaining the original size. Additionally, our proposed model enables the generation of multiple target modalities from a single source modality. We achieve this by the proposed multiple-switchable block, noted as MS-SPADE block , which dynamically transforms source latents to target-like latents according to style. By utilizing conditioning in LDM, we can generate the desired target modality more faithfully. The pipeline of our model is illustrated in Figure \ref{fig1}.

Our main contributions are summarized as follows:
\begin{itemize}[nolistsep]
  \item[•] We propose a model based on the LDM that translates a single source modality to various target modalities (one-to-many) in 3D medical images.
  \item[•] We introduce a switchable block that transfers the source latents to target-like latents using style transfer to enhance the performance of image-to-image translation.
  \item[•] \jhk{}{We validated our method on the BraTS2021 and IXI datasets and despite our model's ability to generate multiple modalities, it demonstrated the highest image translation performance on both datasets.}
\end{itemize}

\section{Related Works}

 \noindent \textbf{Generative adversarial networks} have achieved great success in image-to-image translation by performing continuous adversarial training between the generator and discriminator \cite{NIPS2014_5ca3e9b1, isola2017image, zhu2017unpaired, chen2020reusing, NEURIPS2021_0f281810}. The generator aims to generate as realistic images as possible, while the discriminator learns to distinguish between the generated images and real images \cite{NIPS2014_5ca3e9b1}. For instance, Pix2Pix \cite{isola2017image} focuses on pixel-to-pixel image synthesis based on paired data enhancing the pixel-to-pixel similarity between real and synthesized images. CycleGAN \cite{zhu2017unpaired} is a generalized conditional GAN enabling image synthesis even with unpaired data. NICEGAN \cite{chen2020reusing} proposes a more compact architecture by sharing the encoder parts of the generator and discriminator. compared to conventional GAN. RegGAN \cite{NEURIPS2021_0f281810} adds a registration network to perform image-to-image translation while maintaining anatomical structures in medical images. \jhk{}{Ea-GAN \cite{yu2019ea} is a 3D-based network designed to enhance cross-modality MR image synthesis by integrating edge information, capturing both voxel-wise intensity and image structure details}. ResViT \cite{dalmaz2022resvit} is an adversarial network that incorporates a vision transformer architecture to capture contextual features in medical image translation tasks. However, these methods all work well in situations where there is a one-to-one mapping between the source and target modalities. To generate multiple target modalities from a single source modality, a separate  model is needed for each target modality resulting in an increased number of models.

\begin{figure*} [ht]
    \includegraphics[width=1\textwidth]{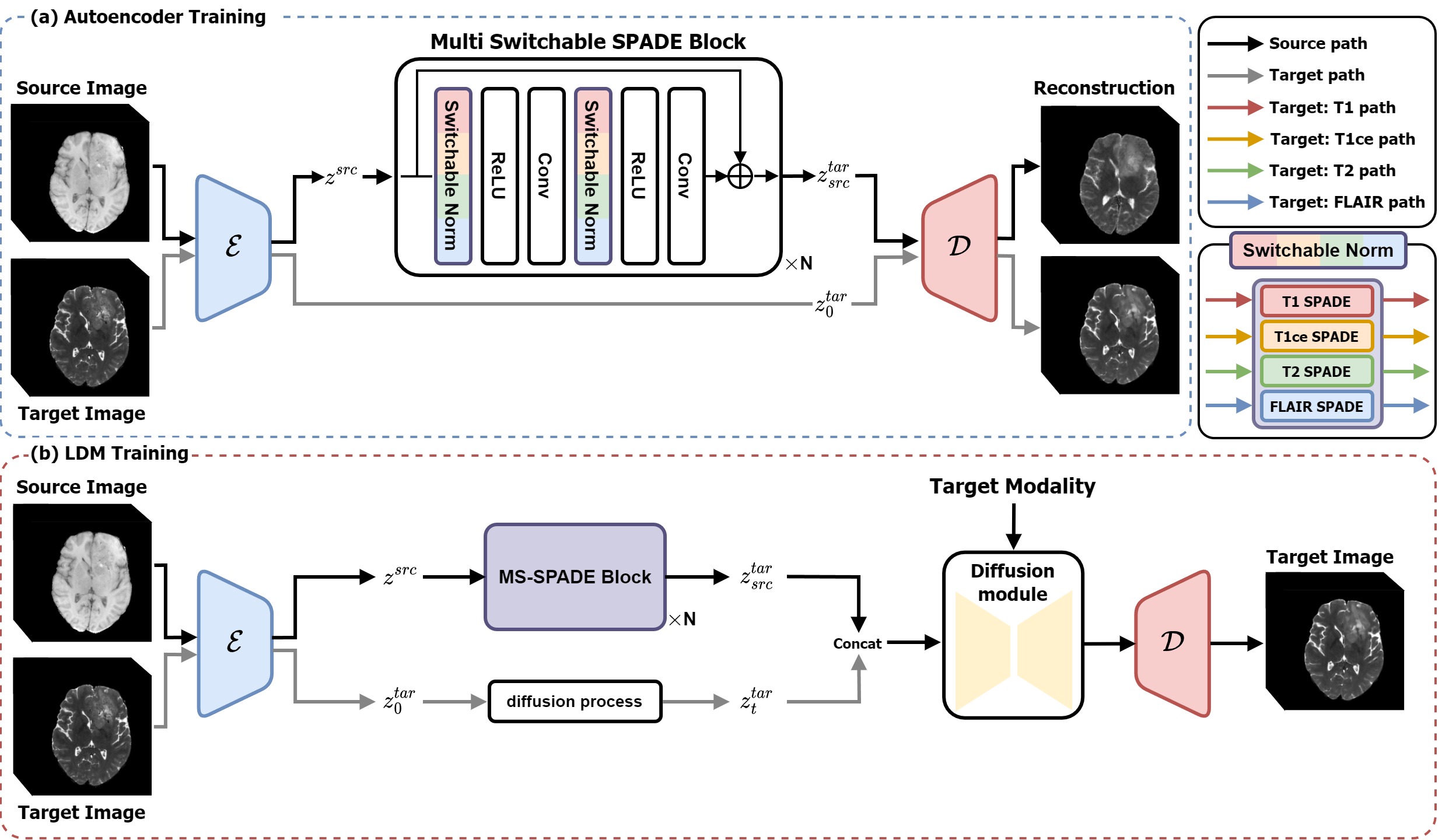}
    \vspace{-6pt}
    \caption{\jhk{}{The source and target images are 3D volume images and our method is applied in a 3D  manner. } In (a), we depict the autoencoder to compute source latents ($z^{src}$ ) and the SPADE block to dynamically convert source latents to target-like latents ($z^{src} \rightarrow z^{tar}_{src}$). The switchable block includes normalization layers that are applied differently depending on the target modality allowing translation to target multiple modalities within a single model. The output of the SPADE block is noted as target-like to emphasize the subsequent role of LDM to fully predict the target latents. The block is stacked N times. (b) illustrates the training process of the LDM model to predict the target latents $z^{tar}$ from the target-like latents $z_{src}^{tar}$ obtained from the SPADE block.}
    \label{fig2}
\end{figure*}

\noindent \textbf{Conditional normalization layers} are techniques used for style transfer and image synthesis tasks \cite{dumoulin2016learned, huang2017arbitrary, perez2018film, park2019semantic}. Among them, adaptive instance normalization (AdaIN) \cite{huang2017arbitrary} was initially used for style transfer tasks and has since been adopted in various computer vision tasks. Unlike other normalization techniques, conditional normalization layers involve normalizing layer activations to zero mean and unit variance, followed by denormalization through learned affine transformations from external data.
Spatially adaptive normalization (SPADE) \cite{park2019semantic} has achieved successful image synthesis by applying spatially varying affine transformations to semantic maps. Unlike previous AdaIN, SPADE generates images that contain both style and semantic information allowing the creation of meaningful images using semantic maps, which is well applicable to 3D multi-modal medical imaging. Inspired by previous research, we adopt the SPADE module that transforms the style and semantic information of an image. Additionally, we propose the multi-switchable SPADE (MS-SPADE) module, which allows for flexible translation to various target modalities.

\begin{figure*} [ht]
    \includegraphics[width=1\textwidth]{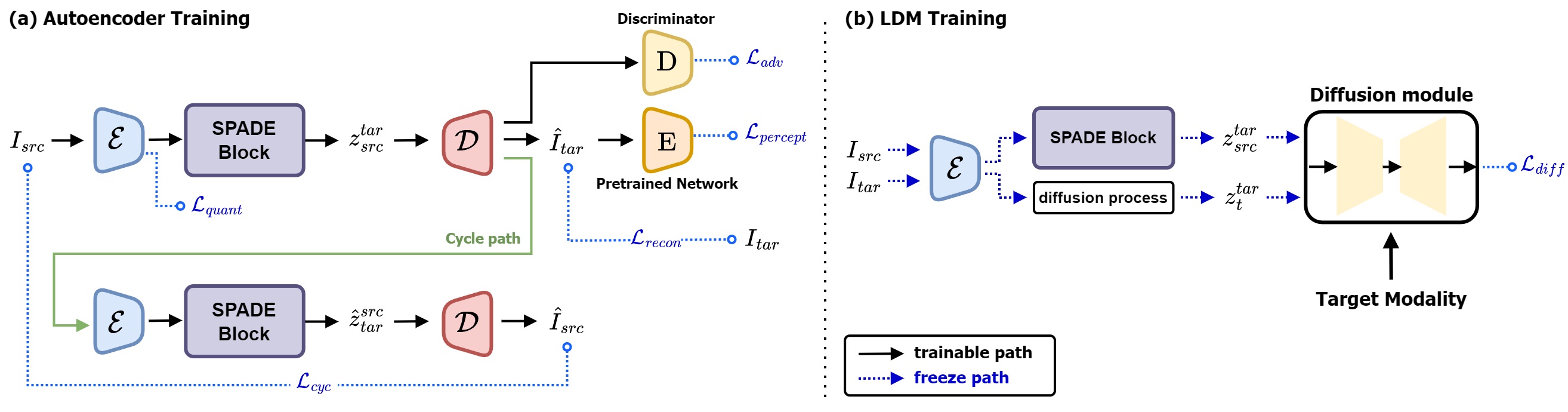}
    \vspace{-12pt}
    \caption{(a) illustrates the training process of the autoencoder during the image compression phase to compute source latents, which involves five different losses: reconstruction, quantization, adversarial objective, perceptual, and cycle consistency. (b) shows the training process of the diffusion model, where the autoencoder is frozen, and only the UNet of the diffusion model is trained. The forward diffusion process is fixed and the input to the UNet consists of the concatenated two latents.} 
    \label{fig3}
    \vspace{-10pt}
\end{figure*}

\noindent \textbf{Diffusion probabilistic model} has recently gained prominence in image generation delivering state-of-the-art results \cite{ho2020denoising, choi2021ilvr, dhariwal2021diffusion, saharia2022palette, pinaya2022brain, croitoru2023diffusion}. Specifically, the denoising diffusion probabilistic model (DDPM) \cite{ho2020denoising} with score matching has demonstrated superior performance in image generation. DDPM leverages a learned Markov chain to transform Gaussian noise distribution into the target distribution. Palette \cite{saharia2022palette} introduces a model that utilizes DM by concatenating source and target images enabling various image-to-image translation tasks such as inpainting, uncropping, image restoration, and colorization without the need for architecture modifications. This model achieves high-fidelity results across different tasks.  LDM \cite{rombach2022high} separates the training process into two stages for efficient processing. First, an autoencoder is trained to provide a perceptually equivalent lower-dimensional representation (e.g.,source latents). Subsequently, DM is trained in this latent space reducing complexity and enabling efficient image generation. Importantly, this approach allows for efficient training without the need for patch cropping in 3D images.

\section{Method}
To overcome the limitations of single-stage generation approaches in recent generative models, numerous studies \cite{razavi2019generating, daidiagnosing, esser2021taming, ramesh2021zero, yan2021videogpt} have adopted a two-stage approach to combine the strengths of different methods resulting in the development of models with superior performance. Our method trains image-to-image translation in a two-stage process as depicted in Figure \ref{fig2}. In the first stage, we train an autoencoder, inspired by previous research methods such as VQ-VAEs and VQGAN \cite{van2017neural, razavi2019generating, esser2021taming}, to compress the image into a latent representation that captures its essential features. The latent space, which captures the perceptual representation of the original images, is regularized via vector quantization and allows for style transfer through conditional normalization layers \cite{park2019semantic}. We also train SPADE block to transform the style of the latents into those of the target modality (as shown in Figure \ref{fig2} (a)). \jhk{}{The proposed MS-SPADE block learns the style parameter of the target and transfers the style from the source to the target.} While translation is possible using only the SPADE module, there is a noticeable difference between the true target latents and those obtained through SPADE and thus we note the output of SPADE as target-like latents. To address this discrepancy, we utilize the LDM to reduce the gap via denoising between these two latents, which is the key idea in our model. Subsequently, we combine the target-like latents (out of MS-SPADE) and noisy target latents (out of forward diffusion), similar to the palette approach, to train the DM (as shown in Figure \ref{fig2} (b)). Our model prioritizes minimizing changes in the source image by first adjusting its latent style and then focusing on generating target-specific latents. 

\subsection{Image compression to compute latents}
Similar to previous work \cite{esser2021taming, rombach2022high, pinaya2022brain}, we employ a perceptual compression model to facilitate the application of the diffusion model in the latent space. We also introduce the MS-SPADE block in the latent space to convert the source latents to the target-like latents. The switchable normalization is illustrated in Figure \ref{fig2} (a). \jhk{}{Our SPADE block learns the mean and standard deviation parameters of the target during training. Thus, during inference, it does not require the target as input but instead uses the target's style parameters for normalization.}

The MS-SPADE block performs different normalizations according to the target modality enabling the transformation of the transformed latents to follow the desired distribution. Let $h \in \mathbb{R}^{N\times C\times H\times W\times D}$ be the input of SPADE Block, where $N, C, H, W$ and $D$ are the size of batch, channel, height, width and depth respectively. The SPADE can be applied as follows:
\begin{equation}  \label{eq:1}
x_{n,c,h,w,d} = \gamma_{c,h,w,d}^{tar}(h)\frac{h_{n,c,h,w,d}-\mu_{c}}{\sigma_{c}} + \beta_{c,h,w,d}^{tar}(h)
\end{equation}

where $x_{n,c,h,w,d}$ denotes the $n,c,h,w,d$-th element of the feature tensor h, $\mu_{c}$ and $\sigma_{c}$ are the mean and standard deviation of the latent features in channel $c$, and $\gamma_{c,h,w,d}^{tar}$ and $\beta_{c,h,w,d}^{tar}$  are modulation parameters that are learned and applied differently according to the target modality during training.

\begin{figure*} [ht]
    \centerline{\includegraphics[scale=0.235]{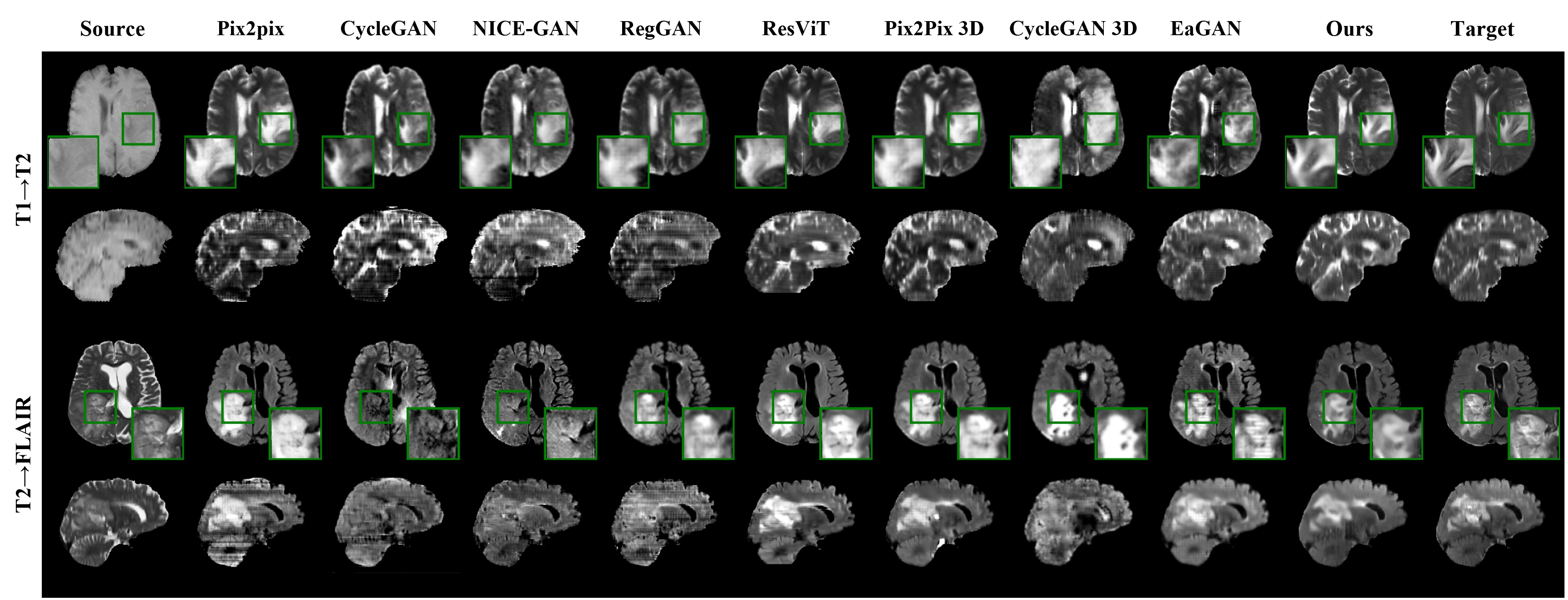}}
    \caption{The figures illustrate the results of the proposed model and comparison models on the BraTS2021 dataset for the image-to-image translation tasks (top: T1 $\rightarrow$ T2, bottom: T2 $\rightarrow$ FLAIR). These two tasks are representative tasks out of 12 possible tasks. The results are depicted in both the axial (top) and sagittal (bottom) views of the 3D volume. The green bounding boxes highlight the important tumor regions in the synthesized images providing a detailed visual representation.}
    \label{fig4}
    \vspace{-8pt}
\end{figure*}

Furthermore, our compression model is based on VQGAN \cite{esser2021taming} and is trained using reconstruction loss, quantization loss, perceptual loss based on previous research and patch-based adversarial objectives \cite{Kingma2014, van2017neural, esser2021taming}. Additionally, we incorporate cycle consistency loss to ensure that the original input can be reconstructed back to the original image through the cycle path \cite{zhu2017unpaired}. The training process is illustrated in Figure \ref{fig3} (a).

\subsection{Diffusion Model}
LDM achieves high-resolution image synthesis while reducing computational cost by training the diffusion process in the compressed latent space \cite{rombach2022high}. Additionally, Palette demonstrates high-quality image synthesis in tasks such as colorization and inpainting reconstruction without relying on conditioning \cite{saharia2022palette}. Instead, it simply takes the target image as an additional input to the DM resulting in superior results. We employ trained perceptual compression models, which consist of an encoder $\mathcal{E}$ and a decoder $\mathcal{D}$, to apply the diffusion model in the latent space \cite{rombach2022high}. Inspired by previous studies \cite{rombach2022high, saharia2022palette}, we design an LDM that takes the target-like latents  besides noisy target latents as inputs. This process is illustrated in Figure \ref{fig2} (b). The latent $z^{tar}_{src}$ represents the source latents transformed to be similar to the target latents by the SPADE block. It is denoted as target-like latents to emphasize the gap between true target latents and those from SPADE. $z^{tar}_t$ denotes the noisy input latents at time step t. Our diffusion model takes $z^{tar}_{src}$ and $z^{tar}_t$ as inputs at time step t to perform the task of predicting the noise $\epsilon$ at that time step. During the training process, the autoencoder used for image compression is frozen, and only the diffusion model is trained (as shown in Figure \ref{fig3} (b)). The loss function is defined as follows:
\begin{equation}  \label{eq:2}
\mathcal{L}_{diff} := \mathbb{E}_{\mathcal{E}(x), \epsilon \sim \mathcal{N}(0,1), t} \Big[ || \epsilon-\epsilon_\theta(z^{tar}_{src}, z^{tar}_t,t)||^2_2 \Big]
\end{equation}

\noindent Our model's neural backbone, denoted as $\epsilon_{\theta}(z_{src}^{tar}, z_{t}^{tar}, t)$, is implemented as a time-conditioned UNet \cite{ho2020denoising ,rombach2022high}. The forward diffusion process is fixed and $z_{src}^{tar}$ and $z_t^{tar}$ can be efficiently obtained from $\mathcal{E}$ during training.

\begin{figure*} [ht]
    \centerline{\includegraphics[scale=0.235]{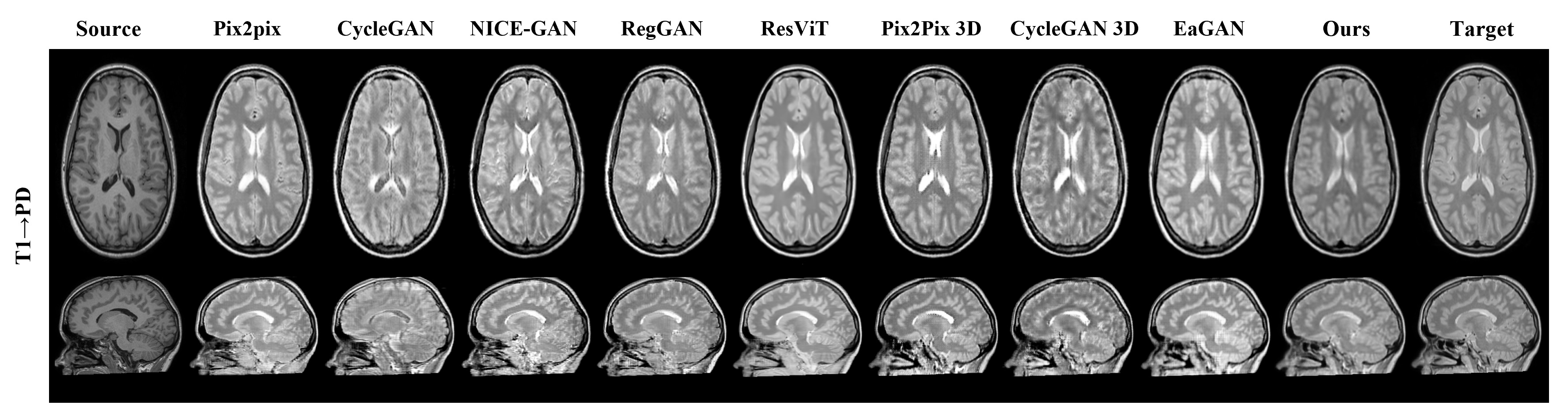}}
    \caption{The figures illustrate the results of the proposed model and comparison models on the IXI dataset.}
    \label{fig5}
    \vspace{-3pt}
\end{figure*}

\begin{table*} [ht]
    \centering
    \scalebox{0.75}{
        \begin{tabular}{ccccccccccc}
            \toprule
            \multicolumn{2}{c}{Datasets} & \multicolumn{6}{c}{BraTS 2021 \cite{baid2021rsna}} & \multicolumn{3}{c}{IXI \cite{ixi2018}} \\
            \cmidrule(lr){1-2} \cmidrule(lr){3-8} \cmidrule(l){9-11}
             \multicolumn{2}{c}{Tasks} & \multicolumn{3}{c}{T1 $\rightarrow$ T2} & \multicolumn{3}{c}{T2 $\rightarrow$ FLAIR} & \multicolumn{3}{c}{T1 $\rightarrow$ PD}\\
            \cmidrule(lr){1-2} \cmidrule(lr){3-5} \cmidrule(l){6-8} \cmidrule(l){9-11}
             \small{Dimension} & Model & PSNR $\uparrow$ & NMSE $\downarrow$ & SSIM $\uparrow$ & PSNR $\uparrow$ & NMSE $\downarrow$ & SSIM $\uparrow$ & PSNR $\uparrow$ & NMSE $\downarrow$ & SSIM $\uparrow$ \\
            \midrule \multirow{5}{*}{2D} 
             & Pix2Pix & 24.624\footnotesize{$\pm$0.962} & 0.109\footnotesize{$\pm$0.028} & 0.874\footnotesize{$\pm$0.015} & 24.361\footnotesize{$\pm$1.061} & 0.117\footnotesize{$\pm$0.021} & 0.846\footnotesize{$\pm$0.019} & 25.283\footnotesize{$\pm$0.861} & 0.095\footnotesize{$\pm$0.028} & 0.882\footnotesize{$\pm$0.023}   \\
            
             & CycleGAN  & 23.535\footnotesize{$\pm$1.334} & 0.155\footnotesize{$\pm$0.035} & 0.837\footnotesize{$\pm$0.028} & 23.418\footnotesize{$\pm$0.944} & 0.164\footnotesize{$\pm$0.033} & 0.825\footnotesize{$\pm$0.035} & 24.854\footnotesize{$\pm$1.128} & 0.106\footnotesize{$\pm$0.033} & 0.841\footnotesize{$\pm$0.027}   \\

             & NICEGAN & 23.721\footnotesize{$\pm$1.136} & 0.148\footnotesize{$\pm$0.029} & 0.840\footnotesize{$\pm$0.029} & 23.643\footnotesize{$\pm$1.045} & 0.148\footnotesize{$\pm$0.022} & 0.829\footnotesize{$\pm$0.033} & 25.330\footnotesize{$\pm$0.939} & 0.102\footnotesize{$\pm$0.027} & 0.855\footnotesize{$\pm$0.026}   \\

             & RegGAN & 24.884\footnotesize{$\pm$0.991} & 0.094\footnotesize{$\pm$0.024} & 0.881\footnotesize{$\pm$0.017} & 24.576\footnotesize{$\pm$1.073} & 0.112\footnotesize{$\pm$0.022} & 0.852\footnotesize{$\pm$0.028} & 25.849\footnotesize{$\pm$0.921} & 0.087\footnotesize{$\pm$0.026} & 0.898\footnotesize{$\pm$0.018}   \\

             & ResViT & 25.578\footnotesize{$\pm$0.812} & 0.088\footnotesize{$\pm$0.021} & 0.895\footnotesize{$\pm$0.018} & 24.825\footnotesize{$\pm$1.030} & 0.108\footnotesize{$\pm$0.018} & 0.861\footnotesize{$\pm$0.021} & 27.265\footnotesize{$\pm$0.847} & 0.076\footnotesize{$\pm$0.022} & 0.916\footnotesize{$\pm$0.021}   \\
            \midrule \multirow{4}{*}{3D}
             & Pix2Pix & 25.181\footnotesize{$\pm$0.861} & 0.097\footnotesize{$\pm$0.031} & 0.887\footnotesize{$\pm$0.012} & 24.602\footnotesize{$\pm$1.181} & 0.113\footnotesize{$\pm$0.021} & 0.854\footnotesize{$\pm$0.018} & 26.387\footnotesize{$\pm$0.849} & 0.089\footnotesize{$\pm$0.032} & 0.903\footnotesize{$\pm$0.019}   \\

             & CycleGAN & 23.740\footnotesize{$\pm$1.198} & 0.138\footnotesize{$\pm$0.032} & 0.835\footnotesize{$\pm$0.019} & 23.508\footnotesize{$\pm$1.301} & 0.152\footnotesize{$\pm$0.039} & 0.822\footnotesize{$\pm$0.024} & 25.507\footnotesize{$\pm$1.133} & 0.099\footnotesize{$\pm$0.035} & 0.882\footnotesize{$\pm$0.028}   \\

             & EaGAN & 24.884\footnotesize{$\pm$0.991} & 0.094\footnotesize{$\pm$0.024} & 0.881\footnotesize{$\pm$0.017} & 24.576\footnotesize{$\pm$1.073} & 0.112\footnotesize{$\pm$0.022} & 0.852\footnotesize{$\pm$0.028} & 26.849\footnotesize{$\pm$0.836} & 0.082\footnotesize{$\pm$0.021} & 0.912\footnotesize{$\pm$0.022}   \\

             & Ours & \textbf{25.818\footnotesize{$\pm$0.857}} & \textbf{0.079\footnotesize{$\pm$0.016}} & \textbf{0.904\footnotesize{$\pm$0.012}} & \textbf{25.074\footnotesize{$\pm$1.085}} & \textbf{0.098\footnotesize{$\pm$0.021}} & \textbf{0.867\footnotesize{$\pm$0.018}} & \textbf{27.729\footnotesize{$\pm$0.885}} & \textbf{0.068\footnotesize{$\pm$0.022}} & \textbf{0.921\footnotesize{$\pm$0.018}}  \\

            \bottomrule
        \end{tabular}
    }
    \vspace{-3pt}
    \caption{ The values represent the quantitative evaluation results of various comparison models and the proposed model in three quantitative metrics (PSNR, NMSE, and SSIM). The reported format is mean followed by standard deviation. The evaluation metrics were calculated on the 3D volume for both the 2D and 3D methods. Boldface entries indicate the top-performing model for each task.}
    \label{table1}
    \vspace{-6pt}
\end{table*}

\subsection{Modality Conditioning}
Inspired by the successful utilization of cross-attention as a conditional image generator in previous studies \cite{choi2021ilvr, rombach2022high}, we incorporate modality conditioning into our model to enhance the image translation to the target modality. Our approach involves converting the given modality into a one-hot vector $y$ and using it as input for the key and value components of cross-attention during training. The cross-attention in the UNet operates as follows:
$\mathrm{Attention}(Q, K, V) = \mathrm{softmax}(\frac{QK^{\mathrm{T}}}{\sqrt{d}}) \cdot V$
with \begin{equation} 
Q = W_{Q}^{(i)} \cdot \varphi_{i}(z_{t}), K = W_{K}^{(i)} \cdot y, V = W_{V}^{(i)} \cdot y
\end{equation}

\noindent Here, $\varphi_{i}(z_{t}) \in \mathbb{R}^{N\times d_{\epsilon}}$ denotes a intermediate representation of the UNet implementing $\epsilon_{\theta}$ and $W_{Q}^{(i)} \in \mathbb{R}^{d\times d_{\epsilon}}, W_{K}^{(i)} \in \mathbb{R}^{d\times d_{y}}$ and $W_{V}^{(i)} \in \mathbb{R}^{d\times d_{y}}$ are learnable projection matrices \cite{jaegle2021perceiver, rombach2022high, vaswani2017attention}. We redefine the loss function with modality conditioning, expressed as $\mathcal{L}_{diff}$ with y representing the one-hot vector for the given modality class.
\begin{equation}  \label{eq:2}
\mathcal{L}_{diff} := \mathbb{E}_{\mathcal{E}(x), \epsilon \sim \mathcal{N}(0,1), t} \Big[ || \epsilon-\epsilon_\theta(z^{tar}_{src}, z^{tar}_t,t,y)||^2_2 \Big]
\end{equation}

\jhk{}{\noindent For the inference process, we utilize the encoder acquired from the image compression phase. We obtain a target-similar latent by passing the source image through the encoder and the MS-SPADE block. This latent is then concatenated with random noise and used as the conditioning for the diffusion model, facilitating the image translation into the target modality.}

\section{Experiments}
\subsection{Datasets}
To evaluate the effectiveness of our model, we utilized the multi-modal brain tumor segmentation challenge 2021 (BraTS 2021) dataset \cite{baid2021rsna, bakas2017segmentation, bakas2017advancing, menze2014multimodal}. We trained our model using the BraTS 2021 training dataset, which consists of 1251 subjects and includes four MRI modalities (T1, T1ce, T2, FLAIR). For evaluating the image translation capability of our model, we employed the BraTS 2021 validation dataset, which comprises 219 subjects. \jhk{}{We also validated our model on the IXI dataset  \cite{ixi2018}, which includes T1, T2, and PD modalities. Out of 574 subjects, 459 were used for training and 115 for testing. For details on the dataset, please refer to the supplementary.}

\subsection{Implementation Details}
The image compression in our model was implemented using VQGAN \cite{esser2021taming} and the diffusion model was based on LDM \cite{rombach2022high}. The implementation was done using PyTorch\footnote{\url{https://pytorch.org/}} and MONAI\footnote{\url{https://monai.io/}} libraries. For detailed information on the model architecture, refer to the supplementary materials. As for the hyperparameters, we used the AdamW \cite{loshchilov2018decoupled} optimizer with an learning rate of $2\times10^{-6}$. 
we set the number of time steps T=1000 with scaled-linearly scheduled noise levels ranging from 0.0015 to 0.0195. The training was conducted on four A100 80GB GPUs with a batch size of 1 per GPU. Additional information on the training hyperparameters can be found in the supplementary materials.

\begin{figure*} [ht]
    \centerline{\includegraphics[scale=0.17]{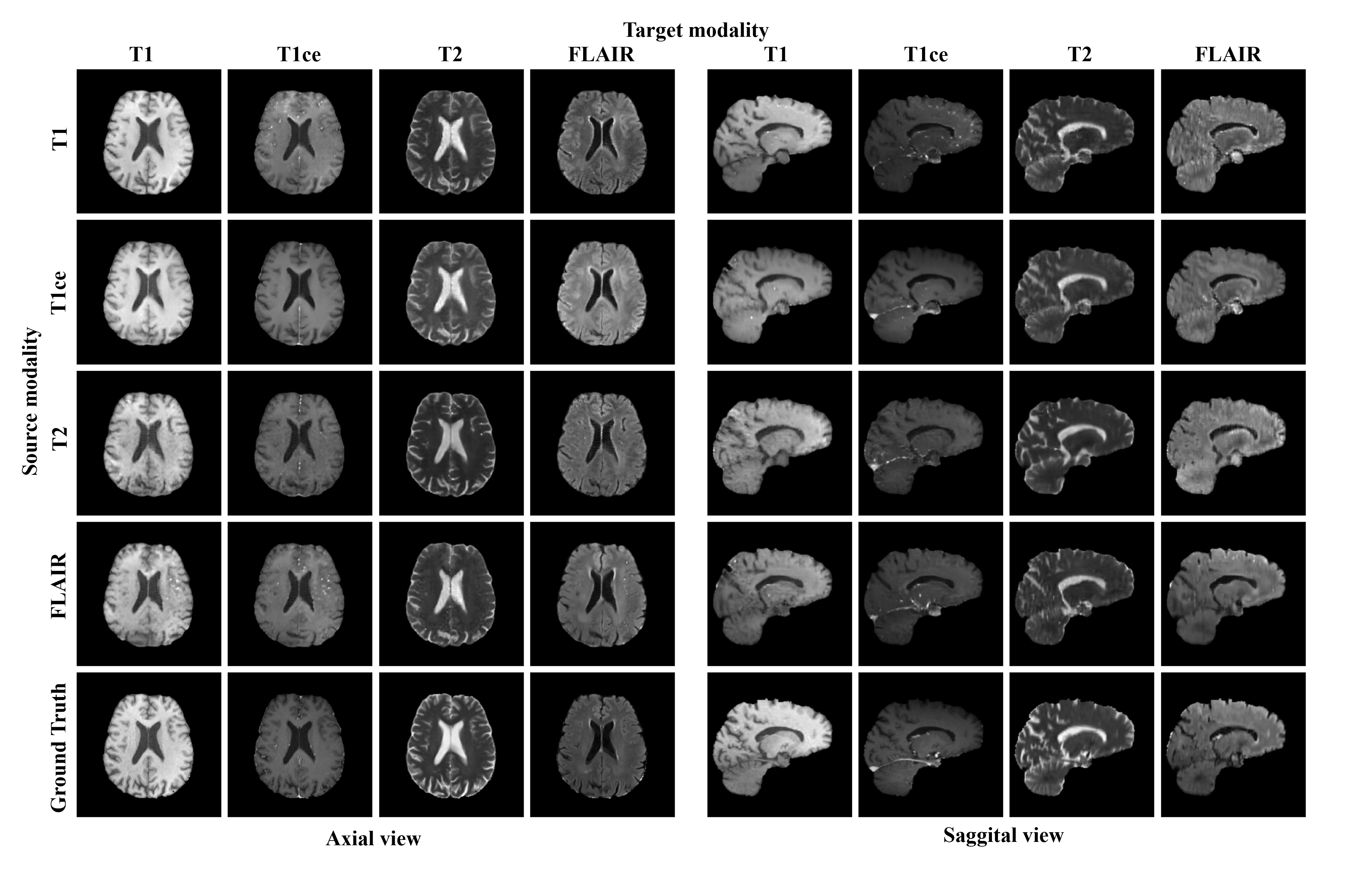}}
    \vspace{-10pt}
    \caption{The figures showcase the image translation results from each source modality to the corresponding target modality using our proposed model for all possible combinations. Each row corresponds to a source modality, each column represents a target modality and the last row represents the ground truth.}
    \label{fig6}
\end{figure*}

\begin{table*} [ht]
    \centering
    \scalebox{0.75}{
        \begin{tabular}{ccccccccccccc}
            \toprule
            \diagbox[height=1.5\line]{Source}{Target} & \multicolumn{3}{c}{T1} & \multicolumn{3}{c}{T1ce} & \multicolumn{3}{c}{T2} & \multicolumn{3}{c}{FLAIR} \\
            \cmidrule(lr){1-1} \cmidrule(lr){2-4} \cmidrule(l){5-7} \cmidrule(l){8-10} \cmidrule(l){11-13}
             Metric &\small{PSNR $\uparrow$}&\small{NMSE $\downarrow$}&\small{SSIM $\uparrow$}&\small{PSNR $\uparrow$}&\small{NMSE $\downarrow$}&\small{SSIM $\uparrow$}&\small{PSNR $\uparrow$}&\small{NMSE $\downarrow$}&\small{SSIM $\uparrow$}&\small{PSNR $\uparrow$}&\small{NMSE $\downarrow$}&\small{SSIM $\uparrow$}\\ 
            \midrule
            \multirow{2}{*}{T1} 
            & \textit{29.001} & \textit{0.055} & \textit{0.942}
            & \textbf{26.119} & \textbf{0.078} & \textbf{0.912}
            & 25.818 & 0.103 & 0.904
            & 24.842 & 0.113 & 0.859 \\
            & \textit{\footnotesize{$\pm$0.643}} & \textit{\footnotesize{$\pm$0.025}} & \textit{\footnotesize{$\pm$0.022}} & \textbf{\footnotesize{$\pm$0.816}} & \textbf{\footnotesize{$\pm$0.022}} & \textbf{\footnotesize{$\pm$0.015}} & 
            \footnotesize{$\pm$0.857} & \footnotesize{$\pm$0.030} & \footnotesize{$\pm$0.014} & 
            \footnotesize{$\pm$0.728} & \footnotesize{$\pm$0.034} & \footnotesize{$\pm$0.019}\\
            
            \multirow{2}{*}{T1ce} 
            & \textbf{26.228} & \textbf{0.076} & \textbf{0.922} 
            & \textit{28.759} & \textit{0.060} & \textit{0.937}
            & 25.990 & 0.092 & 0.907 
            & \textbf{25.204} & \textbf{0.092} & \textbf{0.881} \\
            & \textbf{\footnotesize{$\pm$0.794}} & \textbf{\footnotesize{$\pm$0.027}} & \textbf{\footnotesize{$\pm$0.033}} 
            & \textit{\footnotesize{$\pm$0.885}} & \textit{\footnotesize{$\pm$0.019}} & \textit{\footnotesize{$\pm$0.015}} & 
            \footnotesize{$\pm$0.859} & \footnotesize{$\pm$0.032} & \footnotesize{$\pm$0.908} 
            & \textbf{\footnotesize{$\pm$0.811}} & \textbf{\footnotesize{$\pm$0.050}} & \textbf{\footnotesize{$\pm$0.037}}\\

            \multirow{2}{*}{T2} 
            & 25.422 & 0.085 & 0.908 
            & 25.234 & 0.087 & 0.895 
            & \textit{29.230} & \textit{0.048} & \textit{0.942} 
            & 25.074 & 0.098 & 0.867 \\
            & \footnotesize{$\pm$0.852} & \footnotesize{$\pm$0.026} & \footnotesize{$\pm$0.020} 
            & \footnotesize{$\pm$1.152} & \footnotesize{$\pm$0.034} & \footnotesize{$\pm$0.025} 
            & \textit{\footnotesize{$\pm$0.720}} & \textit{\footnotesize{$\pm$0.018}} & \textit{\footnotesize{$\pm$0.915}} 
            & \footnotesize{$\pm$1.085} & \footnotesize{$\pm$0.021} & \footnotesize{$\pm$0.018}\\

            \multirow{2}{*}{FLAIR} 
            & 25.186 & 0.090 & 0.905 
            & 25.899 & 0.094 & 0.906 
            & \textbf{26.146} & \textbf{0.086} & \textbf{0.913}
            & \textit{28.608} & \textit{0.058} & \textit{0.938} \\
            & \footnotesize{$\pm$0.759} & \footnotesize{$\pm$0.028} & \footnotesize{$\pm$0.048} 
            & \footnotesize{$\pm$1.039} & \footnotesize{$\pm$0.025} & \footnotesize{$\pm$0.027} 
            & \textbf{\footnotesize{$\pm$0.636}} & \textbf{\footnotesize{$\pm$0.028}} & \textbf{\footnotesize{$\pm$0.944}} 
            & \textit{\footnotesize{$\pm$0.769}} & \textit{\footnotesize{$\pm$0.025}} & \textit{\footnotesize{$\pm$0.028}}\\
            
            \bottomrule
        \end{tabular}
    }
    \vspace{-3pt}
    \caption{The values present the quantitative evaluation of image translation results from source modalities to target modalities using our proposed model. The diagonal entries denote reconstructing from latents and are presented for completeness.} 
    \vspace{-8pt}
    \label{table2}
\end{table*}

\subsection{Evaluation Metrics}
We compared the performance of methods with previous approaches commonly used in medical image translation research \cite{yi2019generative}. The synthesis quality was evaluated using peak signal-to-noise ratio (PSNR), normalized mean squared error (NMSE), and structural similarity index (SSIM) \cite{yi2019generative}. They were computed between the ground truth images and the synthesized target images. The average and standard deviation of the metrics were reported on an independent test set that did not overlap with the training-validation sets. The evaluation was performed on 3D volumes for all tests. For the 2D methods, the synthesized target images were stacked to form a 3D volume for comparison.

\section{Results}

\subsection{Comparison of Ours to Baselines}
 \jhk{}{Table \ref{table1} presents the quantitative evaluations of each method for three tasks: T1 $\rightarrow$ T2, T2 $\rightarrow$ FLAIR, and T1 $\rightarrow$ PD, across two different datasets}. These tasks are representative tasks because T1 and T2 are routinely available, while the others are less common. Despite other models performing only one-to-one image synthesis tasks, our model is capable of one-to-many image synthesis and it showed superior performance in quantitative evaluations compared to other models. Moreover, in Figure \ref{fig4}, which depicts qualitative evaluations, our model achieved the closest resemblance to the ground truth and synthesized images that closely matched the challenging tumor regions (green bounding box). Additionally, in the sagittal view, while 2D methods exhibited striped patterns due to their slice-wise inference without considering neighboring slices, our 3D method, including our proposed method, overcame this limitation and demonstrated improved results. These results demonstrate that our proposed model not only allows the generation of multiple modalities from a single modality but also outperforms networks designed for a single task. Furthermore, it shows the successful synthesis of 3D volume images without the need for preprocessing steps such as patch cropping. Figure \ref{fig5} shows the evaluations in the IXI dataset.

\begin{figure*} [ht]
    \includegraphics[width=1\textwidth]{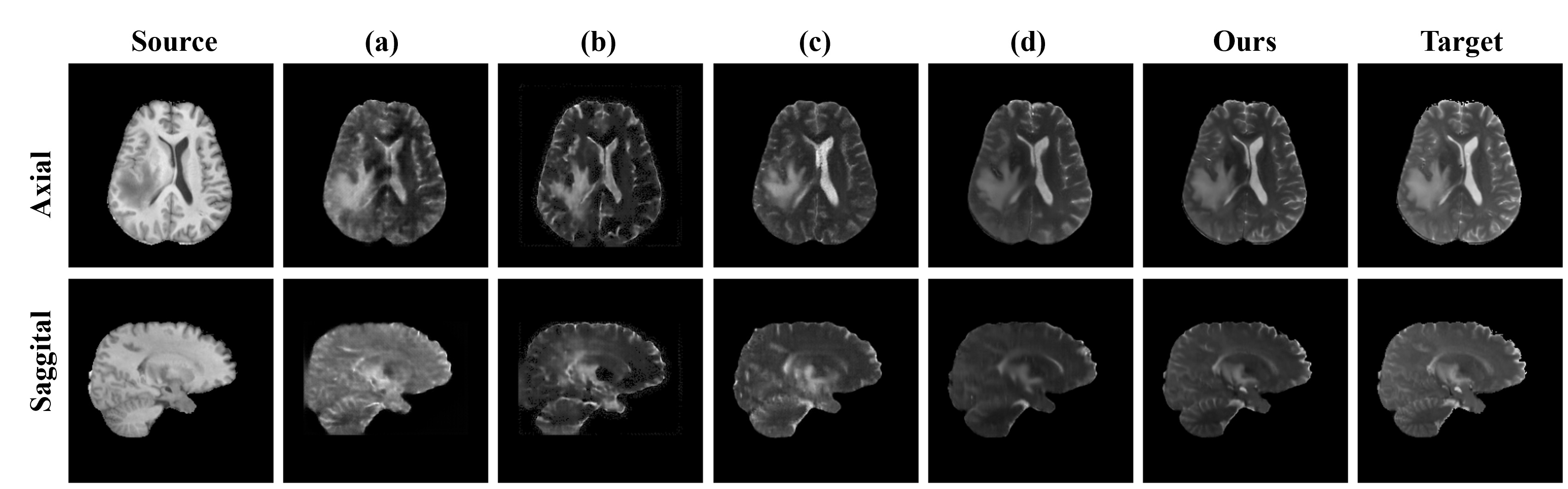}
    \vspace{-12pt}
    \caption{ The figures depict the results of the T1 $\rightarrow$ T2 task for each ablation method on the BraTS2021 dataset. Each method represented in the figure corresponds to the methods listed in Table \ref{table3}}
    \vspace{-8pt}
    \label{fig7}
\end{figure*}

\subsection{Multi Modalities Image Translation}

We conducted image translation experiments using our proposed model with different scenarios of translating one source modality to another target modality given four MRI modalities. Figure \ref{fig6} visualizes the results of image translation tasks from each source modality to the target modality demonstrating successful image synthesis in all cases for the BraTs dataset. The diagonal entries represent tasks where the source and target modalities are the same. The quantitative evaluation of each task in Table \ref{table2} reveals the performance of our proposed model for the BraTs dataset. When the target is T1, the best translation performance is achieved when the source is T1ce. Similarly, for the target of T1ce, the source of F1 yields the highest performance. When the target is T2, using FLAIR as the source produces the best results. Lastly, when the target is FLAIR, the translation performance is highest when the source is T1ce. T1ce, being an MRI with contrast agent administration, contains abundant information about anatomy and tumors compared to other modalities, which explains its superior performance as a source modality in image translation. However, obtaining T1ce, which requires injecting contrast agents can be challenging in many situations. Therefore, our results demonstrate the feasibility of utilizing alternative modalities for image translation when obtaining T1ce is difficult.

\begin{table} [t]
    \centering
    \scalebox{0.6}{
        \begin{tabular}{cccccccc}
            \toprule
             &  &  &  &  & \multicolumn{3}{c}{Metric} \\
            \cmidrule(lr){6-8}
            Method & Diff & Palette & SPADE & Reg-type & PSNR $\uparrow$ & NMSE $\downarrow$ & SSIM $\uparrow$\\
            \midrule
            (a) &\ding{51}&         &         & VQ & 24.902{\small{$\pm$1.073}} & 0.097{\small{$\pm$0.036}} & 0.876{\small{$\pm$0.022}}  \\
            (b) &\ding{51}&\ding{51}&         & VQ & 25.364{\small{$\pm$0.923}} & 0.090{\small{$\pm$0.031}} & 0.893{\small{$\pm$0.024}}  \\
            (c) &         &         &\ding{51}& VQ & 25.037{\small{$\pm$1.019}} & 0.101{\small{$\pm$0.026}} & 0.884{\small{$\pm$0.032}}  \\
            (d) &\ding{51}&\ding{51}&\ding{51}& KL & 25.245{\small{$\pm$0.792}} & 0.095{\small{$\pm$0.022}} & 0.891{\small{$\pm$0.017}}  \\
            ours&\ding{51}&\ding{51}&\ding{51}& VQ & \textbf{25.818{\footnotesize{$\pm$0.857}}} & \textbf{0.079{\small{$\pm$0.016}}} & \textbf{0.904{\small{$\pm$0.012}}}  \\
            \bottomrule
        \end{tabular}
    }
    \caption{The values present the quantitative evaluation metrics of the T1 $\rightarrow$ T2 task for different ablated models: diffusion model, palette, MS-SPADE block, and image compression model for the BraTs dataset. The results are shown and the values in bold indicate the highest performance for each task. Reg-type is the regularization type.} 
    \vspace{-9pt}
    \label{table3}
\end{table}

\subsection{Ablation Study}

To demonstrate the incremental value of design components in our model, we conducted several lines of ablation studies. Specifically, we evaluated the performance contribution of the diffusion model, palette module, MS-SPADE block, and image compression model focusing on the T1$\rightarrow$T2 task using the BraTS dataset. Table \ref{table3} presents the quantitative evaluation results, while Figure \ref{fig7} illustrates the image synthesis results of each method in the ablation study for the BraTs dataset. In Table \ref{table3}, (a) represents the performance of the translation when using the diffusion model alone demonstrating that image translation is achievable. However, (b) with the addition of the palette shows superior performance. Additionally, (c) shows the results of conducting style transfer using only the MS-SPADE block without the diffusion model indicating limitations in image translation performance. (d) and "ours" represent the results with different regularization types of the image compression model applied in previous studies \cite{rombach2022high}. KL-reg, although computationally expensive for 3D training, had an impact on performance due to the adjustment of model size and resulting in slightly blurry compression results. It was observed that the best translation performance was achieved when all the elements were used together.

\section{Conclusion}
\jhk{}{Our method is an important application of computer vision in medicine. We propose a model for multi-modal image translation and conducted comprehensive experiments to evaluate its performance. Our model demonstrated successful image synthesis across different source and target modalities showcasing its versatility. By comparing our method with existing approaches, our method outperformed them in both quantitative and qualitative evaluations.}. Furthermore, our model showcased the ability to perform one-to-many image synthesis, going beyond the limitations of one-to-one tasks performed by other models. We also showed that our model achieved excellent performance even without pre-processing steps such as patch cropping and could successfully perform image synthesis on 3D medical images. \jhk{}{Due to the 3D nature of the proposed approach, our method may be computationally expensive. Future work includes translating from more than two source modalities to a chosen target modality and validating on other medical imaging such as computed tomography.}

\section*{Acknowledgement}
This study was supported by National Research Foundation (NRF-2020M3E5D2A01084892), Institute for Basic Science (IBS-R015-D1), ITRC support program (IITP-2023-2018-0-01798), AI Graduate School Support Program (2019-0-00421), ICT Creative Consilience program (IITP-2023-2020-0-01821), and the Artificial Intelligence Innovation Hub program (2021-0-02068).

{
\small
\bibliographystyle{ieee_fullname}
\bibliography{egbib}
}

\clearpage
\setcounter{page}{1}
\setcounter{table}{0}
\setcounter{figure}{0}
\section*{Supplementary Materials}

\appendix

\renewcommand{\thesection}{\Alph{section}}
\renewcommand{\thefigure}{\Alph{figure}}
\renewcommand{\thetable}{\Alph{table}}

\section{Hyperparameter setting}

In this section, we provide a detailed description of the model architecture and hyperparameters for the autoencoder used in image compression, as well as the structure of the diffusion model. The input dimension for all networks is 3D. For the autoencoder, we applied the network architecture of VQGAN \cite{esser2021taming}, which is explained in Table \ref{tableA}. The structure of the MS-SPADE block present in the bottleneck of the autoencoder is described in Table \ref{tableB}. Additionally, we applied a UNet-based network architecture to the diffusion model used in previous studies \cite{ho2020denoising, rombach2022high}, which is explained in Table \ref{tableC}.

\begin{table}[h]
    \centering
    \scalebox{0.90}{
        \resizebox{\columnwidth}{!}{%
            \begin{tabular}{cccc}
                \toprule
                Input Size & dim $|\mathcal{Z}|$ & Channels & Embedding Size \\
                \midrule
                $192 \times 192 \times 144$ & 8192 & \text{[}256,512,512\text{]} &  3  \\
                \midrule
                Batch Size & Epochs & Model Size & Param Size \\
                \midrule
                1 & 500 & 749M & 237M  \\
                \bottomrule
            \end{tabular}
            }
    }
    \vspace{-6pt}
    \caption{Detailed Hyperparameters for latent diffusion model.} 
    \vspace{-9pt}
    \label{tableA}
\end{table}

\begin{table}[h]
    \centering
    \resizebox{\columnwidth}{!}{%
        \begin{tabular}{cccccccc}
            \toprule
            \multicolumn{8}{c}{MS-SPADE Block} \\
            \midrule
             Stream & Conv. & Act. & Norm. & Conv. & Act. & Norm. & Out ch.\\
             \midrule
             \textbf{In}& $C_{7}$ &  & IN &  & ReLU &  & 128\\
             \midrule
             \textbf{ResBlock}& $C_{3}$ & ReLU & IN & $C_{3}$ & ReLU & IN & \text{[}256,256\text{]}\\
            \midrule
             \textbf{SPADEBlock}& $C_{3}$ & ReLU & MS-SPADE & $C_{3}$ & ReLU & MS-SPADE & \text{[}256,256,256,128\text{]}\\
             \midrule
             \textbf{Out}& $C_{7}$ &  &  &  &  &  & 3\\
            \bottomrule
        \end{tabular}
        }
    \vspace{-6pt}
    \caption{Detailed MS-SPADE Block. $C_{i}$ is the convolution layer with $i \times i$ kernel. $IN$ is the instance normalization layer, and MS-SPADE is the Multi switchable SPADE layer that is applied differently depending on the target modality. Out ch. represents the output channels, and both ResBlocks and SPADEBlocks are repeated 2 and 4 times, respectively} 
    \vspace{-9pt}
    \label{tableB}
\end{table}

\begin{table}[h]
    \centering
    \resizebox{\columnwidth}{!}{%
        \begin{tabular}{ccccc}
            \toprule
            Stream & Condi & Batch Size  & Model Size & Param Size \\
            \midrule
            $48 \times 48 \times 36 \times 3$ &  \text{[}128,256,512\text{]} & 1 & 722M & 658M  \\
            \midrule
            Diffusion steps & Noise Scheculde & $\beta_{start}$ & $\beta_{end}$ & Epochs \\
            \midrule
            1000 & scaled-linear & 0.0015 & 0.0195 & 800 \\
            \bottomrule
        \end{tabular}
        }
    \vspace{-6pt}
    \caption{Detailed hyperparameters for latent diffusion model.} 
    \vspace{-9pt}
    \label{tableC}
\end{table}

\begin{figure*} [ht]
    \centerline{\includegraphics[scale=0.2]{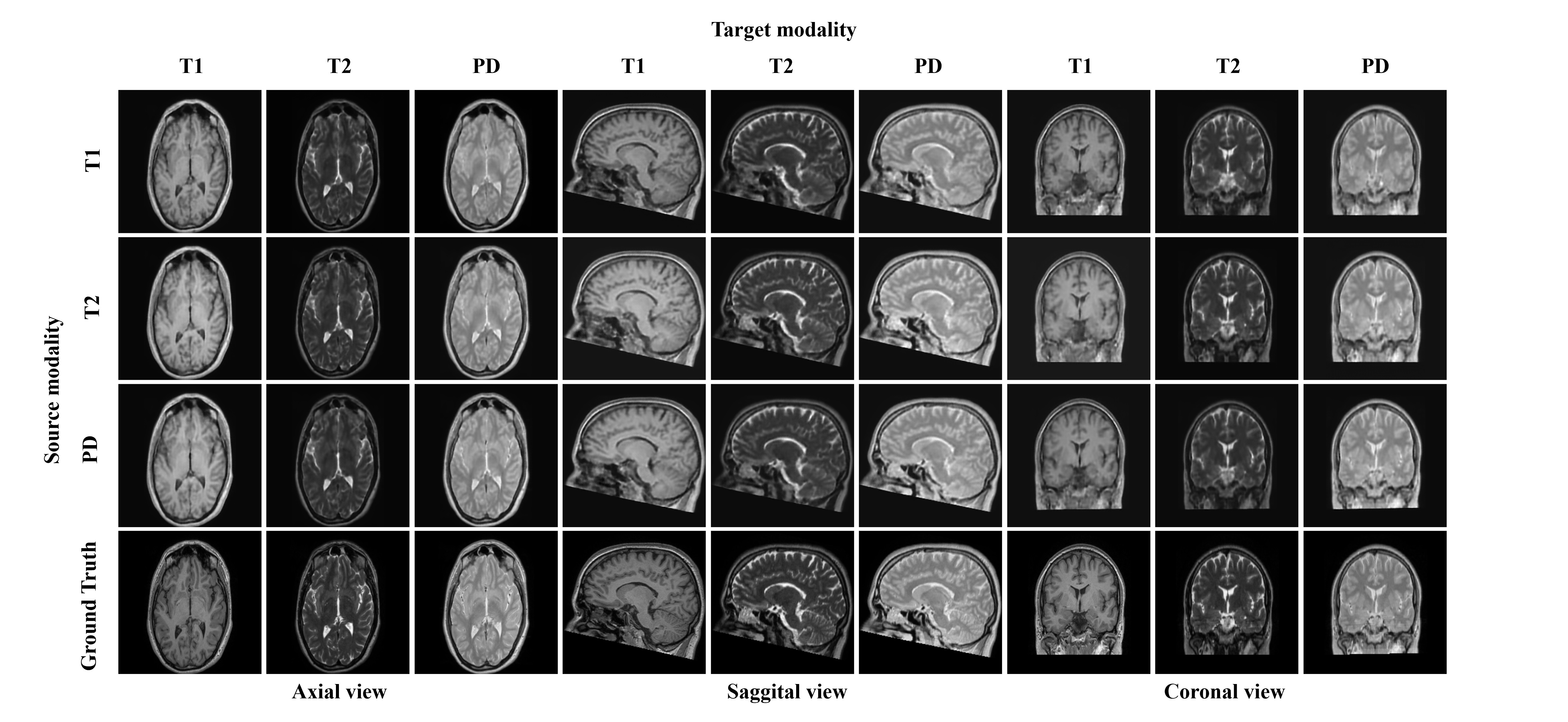}}
    \vspace{-10pt}
    \caption{ The figures showcase the image translation results on the IXI dataset from each source modality to the corresponding target modality using our proposed model for all possible combinations. }
    \label{figA}
\end{figure*}

\begin{table*} [ht]
    \centering
    \scalebox{0.80}{
        \begin{tabular}{cccccccccc}
            \toprule
            \diagbox[height=1.5\line]{Source}{Target} & \multicolumn{3}{c}{T1} & \multicolumn{3}{c}{T2} & \multicolumn{3}{c}{PD} \\
            \cmidrule(lr){1-1} \cmidrule(lr){2-4} \cmidrule(l){5-7} \cmidrule(l){8-10} 
             Metric &\small{PSNR $\uparrow$}&\small{NMSE $\downarrow$}&\small{SSIM $\uparrow$}&\small{PSNR $\uparrow$}&\small{NMSE $\downarrow$}&\small{SSIM $\uparrow$}&\small{PSNR $\uparrow$}&\small{NMSE $\downarrow$}&\small{SSIM $\uparrow$}\\ 
            \midrule
            \multirow{2}{*}{T1} 
            & \textit{29.487} & \textit{0.047} & \textit{0.941}
            & 27.265 & 0.071 & 0.921
            & {27.729} & {0.072} & {0.922} \\
            & \textit{\footnotesize{$\pm$0.522}} & \textit{\footnotesize{$\pm$0.020}} & \textit{\footnotesize{$\pm$0.024}} & {\footnotesize{$\pm$0.629}} & {\footnotesize{$\pm$0.022}} & {\footnotesize{$\pm$0.015}} & 
            \footnotesize{$\pm$0.685} & \footnotesize{$\pm$0.025} & \footnotesize{$\pm$0.018} \\
            
            \multirow{2}{*}{T2} 
            & {27.368} & {0.074} & {0.929} 
            & \textit{29.259} & \textit{0.045} & \textit{0.937}
            & \textbf{27.913} & \textbf{0.067} & \textbf{0.927} \\
            & {\footnotesize{$\pm$0.624}} & {\footnotesize{$\pm$0.031}} & {\footnotesize{$\pm$0.027}} 
            & \textit{\footnotesize{$\pm$0.582}} & \textit{\footnotesize{$\pm$0.017}} & \textit{\footnotesize{$\pm$0.015}} & 
            \textbf{\footnotesize{$\pm$0.659}} & \textbf{\footnotesize{$\pm$0.023}} & \textbf{\footnotesize{$\pm$0.019}} \\

            \multirow{2}{*}{PD} 
            & \textbf{27.968} & \textbf{0.070} & \textbf{0.931} 
            & \textbf{27.834} & \textbf{0.067} & \textbf{0.925} 
            & \textit{29.396} & \textit{0.042} & \textit{0.939} \\
            & \textbf{\footnotesize{$\pm$0.521}} & \textbf{\footnotesize{$\pm$0.028}} & \textbf{\footnotesize{$\pm$0.028}} 
            & \textbf{\footnotesize{$\pm$0.627}} & \textbf{\footnotesize{$\pm$0.024}} & \textbf{\footnotesize{$\pm$0.025}}             & \textit{\footnotesize{$\pm$0.488}} & \textit{\footnotesize{$\pm$0.019}} & \textit{\footnotesize{$\pm$0.027}} \\
            
            \bottomrule
        \end{tabular}
    }
    \vspace{-3pt}
    \caption{The values present the quantitative evaluation of image translation results on the IXI dataset from source modalities to target modalities using our proposed model.} 
    \vspace{-8pt}
    \label{tableD}
\end{table*}

\section{Dataset}
We trained our model on the BraTS 2021 training dataset, encompassing 1251 subjects and four MRI modalities (T1, T1ce, T2, FLAIR). Each MRI scan measures $240\times240\times155$ in dimensions, with a spatial resolution of $1\times1\times1 mm^3$. To assess our model's image translation capabilities, we utilized the BraTS 2021 validation dataset, containing 219 subjects. Additionally, we tested our model using the IXI dataset, including T1, T2, and PD modalities. From the 574 subjects, 459 were allocated for training and 115 for testing. Each of these MRI scans measures $256\times150\times256$ in dimensions with a spatial resolution of $0.9375\times0.9375\times1.2 mm^3$

\section{Comparison Methods details}
To validate the effectiveness of our model, we used commonly used methods in medical image-to-image translation as comparison models. For 2D methods, we employed Pix2Pix \cite{isola2017image}, CycleGAN \cite{zhu2017unpaired}, NICEGAN \cite{chen2020reusing}, RegGAN, \cite{NEURIPS2021_0f281810} and ResViT \cite{dalmaz2022resvit}. For the 3D method, we employed the 3D versions of pix2pix and CycleGAN, as well as the EaGAN proposed as a 3D method, for comparison.
We compared using the discriminator-induced Ea-GAN (dEa-GAN) model as presented in the reference \cite{yu2019ea}. 3D methods are not as commonly used and come with higher computational costs making it challenging to extend existing 2D models to 3D. 
2D methods were executed with a batch size of 32 in the axial view. For the BraTS dataset, they operated on images sized $240\times240$, while for the IXI dataset, zero padding was added to process images at $256\times160$ dimensions. All 3D methods were conducted with a batch size of 1. On the BraTS dataset, images were cropped to $192\times192\times144$ after background removal. For the IXI dataset, images were cropped and padded to measure $256\times160\times224$.

\vspace{1cm}

\section{Additional Experimental Results}
We also analyze which source modality is most effective in synthesizing the target modality within the IXI dataset. Figure \ref{figA} provides the qualitative evaluation results of this multi-modal translation, while Table \ref{tableD} offers the quantitative assessment outcomes. From the qualitative evaluation, we observe that there are minimal differences between modalities, and most present a satisfactory translation performance. As for the quantitative evaluation, it is evident that PD is effective in image translation when generating T1, and similarly for T2. Conversely, T2 proves to be efficient when producing PD images.

\end{document}